\documentclass[preprint]{aastex62}

\graphicspath{{./}{figures/}}

\submitjournal{ApJS}

\shorttitle{Dark data in astronomy}
\shortauthors{Heidorn et al.}

\begin{document}

\title{Astrolabe: Curating, Linking and Computing Astronomy's Dark Data}

\correspondingauthor{P. Bryan Heidorn}
\email{heidorn@email.arizona.edu, gstahlman@email.arizona.edu, julie.steffen@aas.org}

\author[0000-0002-4601-8180]{P. Bryan Heidorn}
\affil{University of Arizona School of Information \\
Harvill Building 4th Floor \\
1103 East 2nd Street \\ 
Tucson, AZ 85721, USA}

\author[0000-0001-8814-863X]{Gretchen R. Stahlman}
\affil{University of Arizona School of Information \\
Harvill Building 4th Floor \\
1103 East 2nd Street \\ 
Tucson, AZ 85721, USA}

\author[0000-0002-8596-6634]{Julie Steffen}
\affil{American Astronomical Society \\
950 N Cherry Ave,
\\ Tucson, AZ 85719, USA}



\begin{abstract}

Where appropriate repositories are not available to support all relevant astronomical data products, data can fall into darkness: unseen and unavailable for future reference and re-use. Some data in this category are legacy or old data, but newer datasets are also often uncurated and could remain “dark”. This paper provides a description of the design motivation and development of Astrolabe, a cyberinfrastructure project that addresses a set of community recommendations for locating and ensuring the long-term curation of dark or otherwise at-risk data and integrated computing. This paper also describes the outcomes of the series of community workshops that informed creation of Astrolabe. According to participants in these workshops, much astronomical dark data currently exist that are not curated elsewhere, as well as software that can only be executed by a few individuals and therefore becomes unusable because of changes in computing platforms.  \deleted{Additional}Astronomical research questions and challenges would be better addressed with integrated data and computational resources that fall outside the scope of existing observatory and space mission projects. As a solution, the design of the Astrolabe system is aimed at developing new resources for management of astronomical data. The project is based in CyVerse cyberinfrastructure technology and is a collaboration between the University of Arizona and the American Astronomical Society. Overall the project aims to support open access to research data by leveraging existing cyberinfrastructure resources and promoting scientific discovery by making potentially-useful data in a computable format broadly available to the astronomical community.

\end{abstract}



\section{Introduction} \label{sec:intro}

Research in astronomy has changed dramatically over the past century: while astronomers 100 years ago would be intimately familiar with a telescope itself, spending many hours observing the sky and making observations in direct contact and collaboration with the instrument, modern astronomers rely heavily on data output, and often on combinations of telescope and archival data \citep{McCray2004,McCray2014}. To successfully answer research questions across astronomical subfields, both old and new data are useful. However, research programs outside of major observatories often lack the infrastructure for proper data management systems to keep older data relevant, and the NSF-funded report \textit{Future Directions for NSF Advanced Computing Infrastructure to Support U.S. Science and Engineering in 2017-2020} points out that an NSF-wide cyberinfrastructure strategy or program does not exist to support disciplinary or cross-disciplinary data sharing and preservation \citep{National2016}. Where appropriate repositories are not available to support all relevant astronomical data products, data can fall into darkness: unseen and unavailable for future reference and re-use. Some data in this category are legacy or old data, but newer datasets are also often uncurated \citep{Hanisch2017} and could remain “dark” \citep{Heidorn2008}.

This paper provides a description of \deleted{the}Astrolabe\deleted{project}, a cyberinfrastructure \added{project} that addresses a set of community recommendations for locating and ensuring the long-term curation of otherwise dark or at-risk data and integrated computing, and in coordination with Open Science initiatives \citep{Open2012}. According to participants in a series of workshops including astronomers, cyberinfrastructure specialists, computational scientists and administrators (see section \ref{sec:workshops}), much astronomical dark data currently exist that are not curated. NASA astronomical missions and large earth-based instruments include planning for data management, and this efficient project-management approach to instrumentation serves the astronomical community well. However, scientists routinely find new ways to analyse, merge and manipulate data, creating new datasets that do not meet the collection requirements of the original projects. This leads to weaknesses in terms of discoverability, compatibility and storage of derivative data. Workshop participants further reported that some software can only be executed by a few individuals and therefore becomes unusable because of changes in computing platforms.  Astronomical research questions and challenges would be better addressed with integrated data and computational resources that span facilities and mission archives. 

As a solution, the design of the Astrolabe system system is based on input from this group of experts, aimed at developing new resources for management of previously-uncurated astronomical data, with an emphasis on small, old and sometimes-heterogeneous datasets. Workshop participants recognized that some newer and large datasets are also uncurated or without long-term data management planning. As described in the sections below, workshop participants provide insight into the data and cyberinfrastructure needs of the astronomical community, including data collection, analysis and publication practices. Recommendations include building on existing astronomy data systems, and borrowing heavily from data and computation systems built for other sciences, with an objective to enable and accelerate new science through improved curation of older as well as newer data. In response to this feedback, we decided to build Astrolabe on the CyVerse cyberinfrastructure\footnote{http://www.cyverse.org}, which was already put into place by the biology community as detailed later in this paper.

Astrolabe collaborates with the American Astronomical Society (AAS), and through this partnership, participants in the Astrolabe workshops suggested that Astrolabe should host data corresponding to published research that are not currently curated by trusted repositories.  The Astrolabe team is also seeking high value datasets as judged by the astronomical community that are not associated with publications. Overall the project aims to support open access to research data by leveraging existing cyberinfrastructure resources and promoting scientific discovery by making potentially-useful data in a computable format broadly available to the astronomical community. The purpose of this paper is twofold: 1) To describe the data management requirements identified by expert participants in a series of workshops for successful curation of dark data in astronomy, and 2) To describe the development of Astrolabe in response to these identified requirements. The paper is outlined as follows: Background information on the dark data problem and open science initiatives is provided, followed by a description of the Astrolabe system and development activities, then presentation of the workshops and outcomes and related discussion, and concluding with a summary and directions for future work.

\section{Dark Data in the Long Tail of Astronomy} \label{sec:darkdata}

Keeping track of dark data in astronomy is important for future scientific discovery. As an illustrative analogy from economics, consider the  emergence of Netflix online while Blockbuster Video store fronts closed around the United States.  Long Tail economics theory \citep{Anderson2004} notes that markets for goods are frequently based on a small number of high volume items. In the age of physical brick-and-mortar stores, the goal is to keep high volume items on the shelves and available for purchase, as low volume items take up space and reduce sales. Best-selling popular movies stocked on Blockbuster shelves are examples of  high volume items, while the many more less popular films and documentaries are low volume items, warehoused by Netflix Online. Netflix was able to economically provide access to low volume items through the Internet, without incurring the costs of physical shelf space in a brick-and-mortar store. Anderson represents this long tail in a graph much like the one depicted in Figure 1 below. The total area under the right part of the curve - the low volume items - represent significant economic opportunity. 

\begin{figure}
  \includegraphics[width=\linewidth]{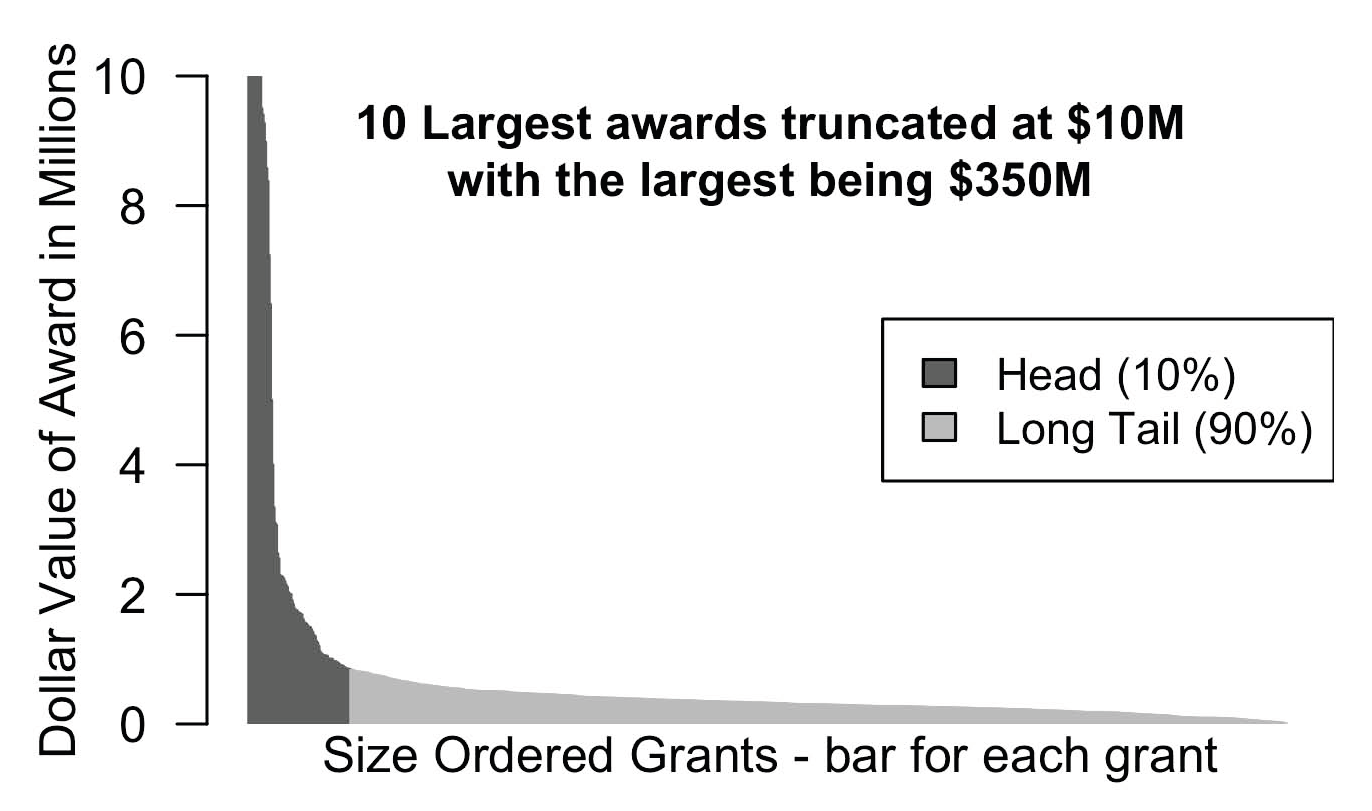}
  \caption{Long Tail of NSF astronomy research funding. The data shown here represents 855 Astronomy and Astrophysics projects. The top 10 percent of projects received between 350,000,000 to about 830,000 dollars. The bottom 90 percent of projects received between about 830,000,000 to 5,000 dollars. The highest-funded projects require a sophisticated data management plan, meaning the Long Tail of research funding likely translates to a long tail of dark (un-curated) data.}
  \label{fig:figure1}
\end{figure}

Long Tail thinking in science was proposed by \citep{Heidorn2008}, who performed an exploration of NSF-funded research grants across disciplines, showing that the largest 20 percent of projects funded in 2007 received more than 50 percent of total funds awarded in that year. While the 2008 paper analysed all of NSF in 2007, Figure \ref{fig:figure1} shows the distribution of funds in NSF Astronomy programs in 2016\footnote{A log transform of this graph would yield a near straight line and thus obscure the tail.}. The astronomy funding distribution is very similar to that of all programs in NSF in 2007 with a few very large projects and many smaller projects. \citet{Heidorn2008} noted that this top 20 percent has well-curated data through access to curation resources and budgeting for data management in engineering and project plans, and that the uncurated data distributed throughout the Long Tail - lower per-year funding of this distribution - represent potentially-valuable information that could be relevant to current research projects and could lead to new discoveries through curation of all data, including data from smaller projects. While there are resources to provide long-term homes for some of these data, much of the data from small projects end up without long term curation. This view is supported by reports of scientists from the workshops discussed in later sections. 

If data are not stored and organized efficiently, scientists will not be able to find and use the data, and these data are essentially “dark”. According to \citet{Heidorn2008}, dark data requires unique curation strategies to capture and manage data that are not easily accessed by potential users. Dark data in the Long Tail are typically: heterogeneous; hand generated; created through unique procedures; curated by individuals; often archived in personal or institutional repositories; not maintained; obscured or protected; seldom reused; and currently unnoticed. This Long Tail data is more likely to be lost over time, in a process referred to in the field of ecology as ``data decay" \citep{Michener1997}.  Without initial good curation, the legacy data in the Long Tail is more likely to be lost than the legacy data from larger and originally well-curated projects. So, both small project size, and associated lack of resources for metadata and curation as well as time (legacy), can combine to threaten data availability. The participants in the workshops presented here largely agreed with this characterization.

\citet{Heidorn2008} states that it is necessary to understand dark data in order to better manage it, a notion that led to and guided the two workshops held in 2015 and 2016 \citep{Stahlman2018}. In the discipline of astronomy, the Long Tail is distinct and multifaceted. \citet{Borgman2015} explains that data volume in astronomy is now growing at a particularly tremendous rate with each new instrument. However, many astronomers implement small research projects, often with specialized instruments for data collection, or rely on derivative and/or theoretical analyses using data available in mission or observatory archives. These data are often not made publicly available, frequently due to the complexity of the datasets themselves, lack of data management skills, and because the utility of data for other researchers may not outweigh the difficulty in providing metadata and stable storage and access \citep{Borgman2015,Edwards2011}. While astronomy is increasingly a ``big data" field, ``little"  or ``small" science – which is the primary focus of the Long Tail theory – is typically hypothesis-driven and led by only one principal investigator rather than a large team \citep{Cragin2010}. Developments in astronomical research towards team-based collaborations and rapid generation of data in astronomy do not necessarily preclude the ongoing presence of “small science” in astronomy, where scientists may generate and hold innovative data in private collections.

\section{Towards Open Data and Open Science} \label{sec:opendata}

Major funding agencies increasingly recognize the importance of public access to research output to facilitate knowledge production, particularly when this research is funded through public support. Many funding agencies now require proposers to provide plans for data management, and also to upload copies of resulting journal articles to public archives. The National Science Foundation \citep{NSF2015}, a leading source of funding for astronomical research and instrument construction in United States, recently published a vision to explore how to improve public access to data, including storage, preservation, discoverability, and reuse, and focusing on data and publications associated with federally funded scientific research. In response to this plan, NSF sponsored a series of workshops within the Directorate for Mathematical and Physical Sciences to obtain feedback from the research community, and to produce recommendations for NSF on realizing this vision. Through this process, the research community is exploring existing initiatives exemplifying best practices and models that could be adapted by NSF \citep{Hanisch2017}, indicating that the topic of open data and supporting cyberinfrastructure is of cutting-edge importance now and for the future of astronomy.

Repositories exist for raw and calibrated astronomy data associated with large facilities, which regularly provide Level 1, 2 and 3 data products.  While there is some variation on definitions, Herschel Data Products\footnote{https://www.cosmos.esa.int/web/herschel/data-products-overview} defines Level 1 as detector readouts calibrated and converted to physical units. Level 2 products are processed for scientific applications and often reviewed by humans. Level 3 are produced from merged level 2 products. Curated repositories include, for example, the National Optical Astronomy Observatory’s DataLab\footnote{http://datalab.noao.edu/}, which connects catalog objects with NOAO images and additional services for data analysis. Other repositories such as the NASA/IPAC Extragalactic Database (NED)\footnote{https://ned.ipac.caltech.edu/} and VizieR\footnote{http://vizier.u-strasbg.fr/viz-bin/VizieR} are trusted collections of catalogs, tables and images. Along with institutional repositories, Harvard's Dataverse\footnote{https://dataverse.harvard.edu/} repository and CERN's Zenodo\footnote{https://zenodo.org/} are currently-used resources for some astronomers with intermediate data products and data associated with publications that are not archived elsewhere. The value of data archives has been demonstrated, for example, by \citet{White2009}, where the number of papers based on archival data from the Hubble Space Telescope exceeded the number of non-archival publications. While much data are preserved and accessible in well-curated repositories as described above, many other data products - including data derived from analyses of mission data, synthesis of data from multiple missions or instruments and supporting published research - often remain hidden and require development of resources for data access and preservation throughout the lifecycles of these data \citep{Conrad2017}. 

The National Virtual Observatory (NVO) initiative was conceptualized in a 2001 white paper \citep{Brunner2001} and evolved into a series of Virtual Observatory (VO) projects and products. The NVO was envisioned as a semantic web of ontologically-linked knowledge encompassing the full research lifecycle of archived raw and derived astronomical data, computation and software, as well as project proposals and publications \citep{Accomazzi2011a,Hanisch2007,Brunner2001}. Furthermore, the NVO was expected to collaborate with academic research libraries for long-term curation \citep{Choudhury2008}. VO technology and standards are now actively used throughout the international astronomical community, representing an opportunity for linking data in the Long Tail to corresponding literature and broadly facilitating discoverability. \replaced{Further}{Integral to} supporting astronomical research, the Astrophysics Data System (ADS) is a robust open-access tool, indexing nearly all astronomical publications and associated metadata, including links to data when available \citep{Accomazzi2011b,Accomazzi2016}, and with capabilities for visualizing citation and collaboration patterns \citep{Henneken2009}. Using the ADS, \citet{Henneken2011} show that publications with links to data are more highly cited than publications that do not link to data, an important finding for open data and data sharing initiatives \citep{Henneken2015}. 	

\section{ASTROLABE}

As shown in previous sections, where appropriate repositories do not exist, data can fall into darkness. A functional repository must make it easy to ingest, format and normalize, index for discoverability, transform, visualize and deliver data for reuse. Fortunately, not all of this functionality must be built from scratch. Astrolabe is a new \deleted{cyberinfrastructure}\added{system} for storage and analysis of previously uncurated astronomical data, built using the existing CyVerse\footnote{cyverse.org} cyberinfrastructure platform. Many of the Astrolabe design elements within CyVerse have been created in direct response to recommendations of scientists in the workshops presented below. As a community-oriented and multidisciplinary cyberinfrastructure designed to support open science, CyVerse collaborates with other repositories to ensure interoperability and ongoing development and deployment of its versatile technology across institutions \citep{Lenhardt2016}. Flexible cyberinfrastructure (i.e. infrastructure based on distributed computer, information and communication technology) is key to these objectives. CyVerse provides a cloud computing environment that helps to free end users from the necessity to install and configure sometimes-complex systems of programs on their laboratory computers. Originally known as the iPlant Collaborative and supported by NSF since 2008 \citep{Atkins2003}, CyVerse provides computation, data storage, portability and data federation (controlled copies) though iRODS\footnote{https://irods.org/} data management software.  Links to supercomputing resources for cloud-based data analysis and software tool development provide scalability. These features have attracted more than 31,000 active users and nearly 3,000 Terabytes of user data \citep{CyVerse2017}. Initially designed for the biological community, CyVerse is expanding its services to other disciplines. Projects from a variety of disciplines across Natural and Life Science domains are now based in CyVerse technology. This type of computation supports software and data reusability, repeatability of computations, and overall validity in science.

Astronomy has produced a rich set of processing resources and standards. However, these tools and standards sometimes require very particular processing environments, making it difficult to develop an integrated operating environment for astronomical data. Astrolabe workshop participants suggested using CyVerse as one of the platforms for delivering astronomical repository and processing services, particularly because the CyVerse cyberinfrastructure elements adapt well to the needs of a variety of scientific disciplines, including astronomy. CyVerse supports Docker\footnote{https://www.docker.com/} technology, which allows developers to package together all of the languages, libraries and other tools and their dependencies and wrap them into portable containers, thus isolating users from the complexity of the relationships. CyVerse also support iRODS to allow for easy control of workflows even for large datasets, as well as movement of data between repositories, supporting long-term sustainability. Furthermore, CyVerse provides access to scalable computing with relatively easy access to high performance computing for end users.

The Virtual Observatory project provides useful guidelines for Astrolabe development, including basic discovery functionality and data federation to allow search across repositories. Generally, raw data generated by an instrument must be processed into a form that is more directly useful for astronomers, and this involves processes such as calibration and registration, as well as maintenance of metadata. There are multiple data formats in astronomy, but a few formats dominate \citep{Greenfield2015}, including Flexible Image Transport System (FITS) \citep{Grosbol1988} and Hierarchical Data Format 5 (HDF 5) \citep{Folk2011}, and Astrolabe is creating workflows to handle processing of these data types. We are also currently directing Astrolabe efforts towards linking tools such as AstroPy, PyRAF, CarPy, JS9 and Montage to CyVerse to provide astronomers with easy access to these applications within the repository environment and through an attractive user portal.

\begin{figure}
  \includegraphics[width=\linewidth]{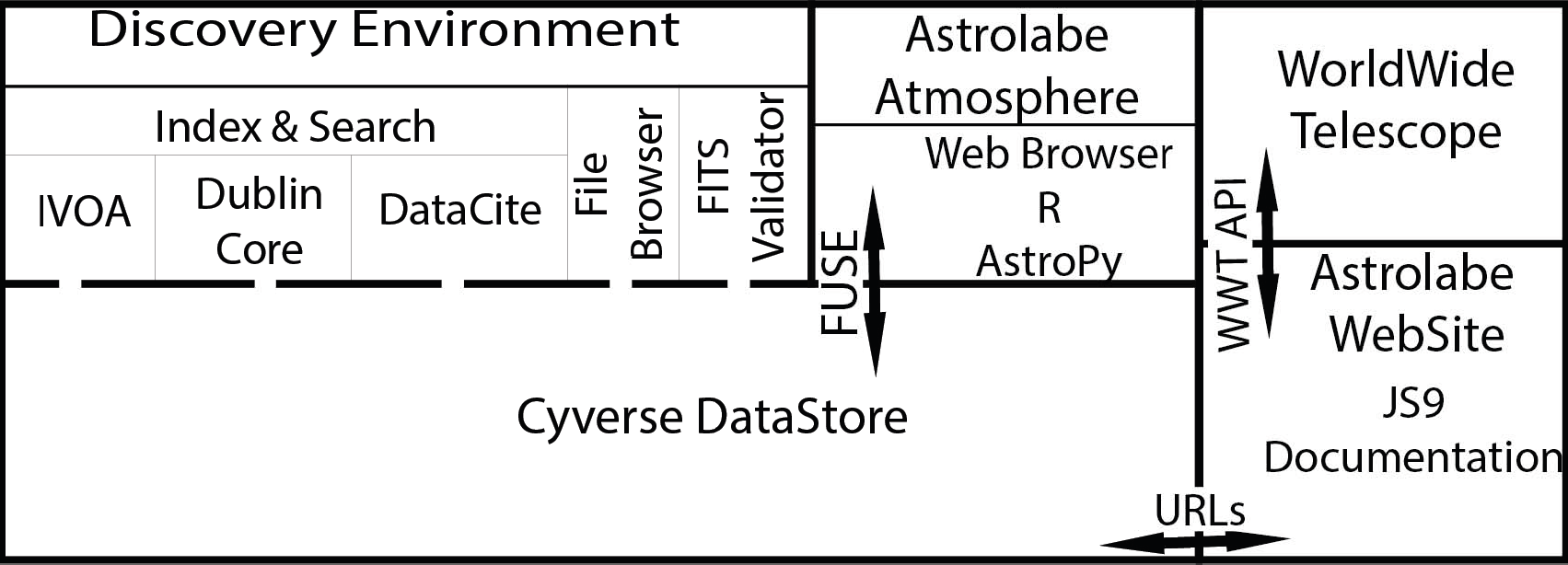}
  \caption{Astrolabe system diagram}
  \label{fig:figure2}
\end{figure}

Below we discuss the Astrolabe/CyVerse ecosystem (depicted in Figure \ref{fig:figure2}), and briefly outline how an astronomer can begin using Astrolabe.

\subsection{The Astrolabe-CyVerse Eco-System}

Astronomers can use Astrolabe to store and share data, and importantly they can process data in a cloud environment. The Astrolabe repository uses CyVerse as a storage and analysis environment for heterogeneous astronomy data, and provides digital object identifiers (DOIs) to link data to corresponding literature and to other established databases for overall transparency and reuse of research outcomes in Physical Sciences. Furthermore, Astrolabe is developing WorldWide Telescope (WWT) open-source software as a scalable repository front-end and tool for visualizing spatial data such as astronomical or planetary images on the sky and in 3-dimensions; see \citet{Rosenfield2018}. Astronomers may interact with Astrolabe through the following established links to the Cyverse Data Store (which holds the actionable data for Astrolabe): 1) Users may enter the Astrolabe website and directly access select items in the Astrolabe repository or use our web services in a stand alone manner (e.g., a JS9 implementation that connects to WorldWide Telescope); or 2) Users may enter through the CyVerse Discovery Environment (DE), to upload data to the Astrolabe repository (using the CyVerse DataStore), search metadata to find items in the repository, and/or run Dockerized applications in the DE. Users can also analyze data in the repository using astronomical software with cloud services (Atmosphere). We now discuss each of these products and services below.

\paragraph{CyVerse DataStore} The CyVerse DataStore is a cloud-based storage system with parallel IO for rapid movement of data. Astrolabe users have fine-grained control of access through authentication services. Astrolabe has multiple terabytes of storage, and each individual Astrolabe user is allocated an additional 100 gigabytes (current transfer rates are about 1 gigabyte per 18 seconds). The DataStore can be accessed directly using industry standard protocols such as iRODS (iCommands)\footnote{https://irods.org} and via batch mode using file transfer tools such as CyberDuck\footnote{https://cyberduck.io}, which supports iRODS. Astrolabe data exists in two main partitions - as a private space, and as public “Community Data” that will be discussed in the following description of the Discovery Environment. 

\paragraph{Discovery Environment} The Discovery Environment (DE) is a web interface that provides access to data and applications that can run on the CyVerse hardware cluster or on Texas Advanced Computing Center (TACC) supercomputers. Data can be accessed via browsing or through search over multiple metadata frameworks with CyVerse ElasticSearch\footnote{https://pods.iplantcollaborative.org/wiki/display/~bjoyce3/ElasticSearch+Integration+in+CyVerse}. Astrolabe currently uses three metadata indexing and search templates: Dublin Core\footnote{http://dublincore.org/} and DataCite\footnote{https://www.datacite.org/} (both are native to CyVerse), as well as a  new metadata template created for Astrolabe following the IVOA ObsCore\footnote{http://ivoa.net/documents/ObsCore/index.html} standard and the FITS and WCS standards\footnote{https://fits.gsfc.nasa.gov/}. The Astrolabe metadata template  can be populated directly from imported FITS files by using a Discovery Environment FITS Metadata Extractor application. The Astrolabe metadata template and the DataCite template each include the capability to select Unified Astronomy Thesaurus \citep{Frey2018} terms as keywords in the subject field. DOIs can be requested after assigning DataCite metadata to datasets and/or files. To ensure the integrity of this process, Astrolabe is also participating in ongoing discussions with other institutions and projects about best practices for assigning DOIs to astronomical data; see \citet{Novacescu2018}. DataCite metadata records and DOIs are only assigned to collections of data that are considered to be permanent and unchanging with appropriately detailed descriptions. Scientists can also create and share applications (apps) in the DE (the FITS Metadata Extractor application is an example). Hundreds of apps currently exist, though most are designed for biology. All DE apps are “containerized” with Docker allowing them to execute on multiple platforms. 

\paragraph{Atmosphere Virtual Machine} An Astrolabe Atmosphere virtual machine has been pre-configured with astronomical data analysis software through CyVerse's “Atmosphere” environment. An Astrolabe user can run an instance of this machine, perform tasks such as editing images with JS9, and then save the outputs conveniently back into the user's Astrolabe CyVerse DataStore account. Astrolabe Atmosphere users can boot servers from 1-16 processors, memory from 4-128 gigabytes of RAM and 30-1200 gigabytes of disk depending on the processing needs. The Atmosphere environment is being deployed in the NSF-funded JetStream project \citep{Stewart2015,Towns2014}. Planning for JetStream compatibility allows for sustainability and potential for migration to an environment that can provide hundreds of processors. Astrolabe's Atmosphere implementations of Astropy and R-Suite are under development. Other astronomical software such as PyRAF/IRAF will be added based on user demand. Astrolabe's Atmosphere runs a web browser that can execute the modified JS9 implementation running on the Astrolabe web server discussed below and could run other web services as well. This allows the DataStore to be accessed directly through JS9. A detailed workflow for using this software is provided below as well.

\paragraph{Astrolabe Web Server} The Astrolabe web server is hosted on the CyVerse Cloud. The website\footnote{http://astrolabe.arizona.edu} hosts information on the project, hyperlinks to select collections in the DataStore, as well as an instance of the JS9 public API implementation\footnote{https://js9.si.edu/js9/help/publicapi.html}.  JS9 can be used to edit FITS metadata and images for visualization in WWT or direct download. JS9 can also create FITS files from other image types, such as a .jpg or .png. Furthermore, this JS9 implementation can run using a web browser in the Astrolabe Atmosphere environment to allow additional processing and direct access to files in the CyVerse DataStore.

Astrolabe does not use all CyVerse functionality, and other services may be added to Astrolabe. The process of accessing and utilizing Astrolabe through CyVerse is discussed in the next section.

\subsection{Using Astrolabe}

As mentioned above, CyVerse was initially developed to support Life Sciences, but has since endeavored to expand to other disciplines in the United States research community. Prior to the inception of Astrolabe, CyVerse was used by only a few individual astronomers for particular instruments and data analysis tasks. Astrolabe now endeavors to develop attractive and useful tools and widely connect astronomers to CyVerse resources, augmenting an already-substantial multidisciplinary user community with astronomical researchers and data. As a first step towards creating a fully-integrated web portal to CyVerse, users and potential-users can learn about Astrolabe features and how to access CyVerse through the project's website: \url{http://astrolabe.arizona.edu}. 

Detailed instructions on using Astrolabe are provided on the Astrolabe website. To summarize, users must first create a CyVerse account at \url{http://www.cyverse.org/} to join the Astrolabe community. \added{The users can then login to the Discovery Environment, create a subdirectory, upload data, set permissions to share data (or keep some private), add metadata for discoverability, edit images or FITS headers and view images in the WorldWide Telescope.}\deleted{then follow these general steps:

\begin{enumerate}
\item Create a subdirectory for data: Login to the Discovery Environment at \url{https://de.cyverse.org/de/}. Select the blue ``Data" icon on the top-left and create a new folder.
\item Upload data: Data can be uploaded by dragging and dropping the files into the new folder. The Astrolabe team can assist with transferring multi-gigabyte datasets and files, and CyVerse provides tutorials for this as well \footnote{http://www.cyverse.org/data-store}.
\item Give Astrolabe access: Give the Astrolabe Project access to your data or make it public to all. From the Discovery Environment, click the ``share" button and share with the Astrolabe PI ``heidorn". 
\item Make data discoverable: Metadata templates can be assigned to index data for searchability and Virtual Observatory compatibility, and to assign permanent DOIs to static datasets. 
\item Edit FITS images using JS9 and open in WWT: The JS9 interface can be conveniently accessed through the Astrolabe website for image modification and automatic visualization in WorldWide Telescope. 
\item Edit FITS header: A tool can edit FITS headers to ensure data quality. To save modified FITS file back into the CyVerse data store, an Atmosphere virtual machine can be used. 
\end{enumerate}}

\paragraph{Pilot Data} Astrolabe is actively seeking users and datasets. At the time of this publication, the project has successfully ingested several datasets, including the AAS CD-ROM Series, which was launched in 1993 and includes nine volumes of data corresponding to publications between 1993-1997. These data are primarily catalogs and tables, and more than 300 folders have been indexed in CyVerse with preliminary DataCite (DOI) metadata, searchable with CyVerse's “beta” search function.

\section{Astrolabe Community Workshops} \label{sec:workshops}

Informing the creation, design, and implementation of Astrolabe, two separate workshops were held for communities of researchers, computational staff and administrators in astronomy. Participants in both workshops were primarily recruited from the University of Arizona, National Optical Astronomy Observatory (NOAO), the AAS and CyVerse. Data collected from the workshops include: field notes taken by workshop organizers using a collaborative note-taking platform; preliminary surveys administered online; audio recordings of presentations, group discussions and breakout sessions; and transcription of key dialogue. This material was analyzed qualitatively by coding for themes and sub-themes. The workshop methods and results of these analyses are summarized below.

\subsection{Preliminary Surveys}

Guiding the design of the two Astrolabe workshops and session topics, an identical short, informal, open-ended online survey was sent to all workshop invitees prior to each event. Questions targeted: technical and social issues associated with uncurated astronomy data; strengths and weaknesses of existing data repositories; science use cases for a new astronomy repository; opportunities for education and outreach; and opportunities and challenges associated with the proposed project. Respondents provided substantial text-based feedback and inquiries, and key themes were identified prior to each workshop and addressed through workshop breakout sessions.

\subsection{Astrolabe Workshop Design}

Invitees to the 2015 ``Arizona Astronomical Data Hub - AADH" (the initial tentative name for the Astrolabe project) workshop included astronomy and data science experts, primarily affiliated with University of Arizona, NOAO and the AAS, with several invited speakers, and remote participation by individuals at the Harvard-Smithsonian Center for Astrophysics. The workshop itself loosely followed the Delphi Method of achieving group consensus \citep{Keeney2001}. Thirty-two participants (remote and in-person) and facilitators were organized into groups based on expertise, ensuring that each group contained a representative mix of astronomers, data scientists, librarians, and educators/administrators. Following an initial introduction by guest speakers, four breakout sessions were held over two days, each immediately followed by whole-group report-out sessions. Detailed notes were taken by each facilitator and by participants using a collaborative note-taking platform, and invited talks and whole-group discussions were recorded; this material was analyzed and coded to identify the following broad themes and detailed objectives for the Astrolabe project.

A follow-on 2016 Astrolabe workshop included twenty-two local and nonlocal participants. The group was comprised of research astronomers, University of Arizona and CyVerse representatives, and other expertise related to astronomy data and publishing, along with five workshop organizers. Just prior to the Astrolabe workshop, the first meeting of the Astrolabe Advisory Board was convened. Board members broadly discussed the research challenges influencing the creation of Astrolabe, as well as the general function and operation of the Board. Along with the preliminary online survey that was distributed to participants, this meeting helped to inform the design of the workshop's three breakout sessions and related group discussions. 

The 2016 workshop began with a welcome and introduction from the AAS. Workshop participants were then updated on Astrolabe, the project's activities and progress over the previous year, as well as the workshop structure and objectives. A series of invited talks provided an overview of the resources available through CyVerse, the UAT, and a general discussion of data management and visualization, and publishing in astronomy, along with a demonstration of the functionality and new searching capabilities of the ADS index of astronomy literature. The first day's afternoon session asked four breakout groups to discuss the most important and practical solutions, data types, and astronomy subfields to direct focus for incorporating dark data that is both useful and accessible into Astrolabe. When the workshop reconvened the following morning, each breakout group briefly discussed the prior day’s conclusions and then reported its findings. The key areas of focus identified by the four breakout groups were listed, and participants voted on the four most critical and practical. Subsequent breakout sessions were organized around four topic areas recommended for the immediate attention of Astrolabe. Workshop participants explored these topics in depth during a breakout session directed towards the functional requirements that would be necessary to address each topic from both a user and developer perspective. The final session was condensed to three breakout groups to discuss funding opportunities and overall strategies for the sustainable growth of Astrolabe. Several additional invited talks on the second afternoon provided further context for Astrolabe through a discussion of the strengths and weaknesses of several case studies. The workshop concluded with a final whole-group discussion.

The key topics identified below were sometimes discussed in more than one of the two surveys or workshops but are presented only once for simplicity and supported by relevant participant feedback. This validation of the dark data problem (and surrounding issues and opportunities) is evident in recorded responses to the pre-workshop survey sent to all participants (8/32 responses in 2015, 7/22 responses in 2016). Survey responses are interspersed to illustrate the qualitative insights derived from the 2015 and 2016 workshops in the following section.

\section{Workshop Outcomes}
\label{sec:outcomes}

\subsection{2015 Astrolabe Workshop Outcomes}

The following broad themes and detailed objectives were identified for the Astrolabe project and confirmed by pre-workshop survey responses:

\subsubsection{Identify mission and clear science use cases}	

One or more detailed science use cases are essential to obtaining support and creating a viable new repository for dark or uncurated data in astronomy. Therefore, establishing and documenting potential science cases are important objectives of this project. Workshop discussions indicated that certain existing survey datasets are in danger of becoming dark data and could illustrate the utility of a new astronomy data repository. For example:

\begin{quote}
Each snapshot of the universe is unique. Data may represent the only image or spectrum corresponding to a particular timestamp in a particular direction, certainly from a particular observatory location. Simply collecting and protecting the data should be the most fundamental goal. (Anonymous Respondent, 2016 pre-workshop survey)
\end{quote}

In addition, it was recommended that the project mine the published literature for references to uncurated datasets. Key among the suggested use cases is the time domain, where new discoveries lead to interest in old observations, which could be dark data:

\begin{quote}
The generic argument for curating every direct image ever taken is the potential value for time domain studies - searching backwards in time for historical outbursts of unstable systems, long-term stellar variability for exoplanet hosts, multiple supernovae in the same galaxy, or coming soon - optical counterparts to gravitational wave signals and/or gamma ray bursts extracted from the multi-messenger data stream after the fact. (Anonymous Respondent, 2016 pre-workshop survey)
\end{quote}

\subsubsection{Social and technical barriers}

Participants acknowledged a number of social and technical barriers to explain why the dark data problem exists. Some barriers and solutions are addressed in subsections of the Discussion presented in this paper. Three commonly referenced barriers encapsulated in the following comment are lack of funding, lack of time, and lack of credit:

\begin{quote}
Lack of interest in dark data by funding agencies. Lack of credit for curation of dark data for employment selection, promotion, or tenure. Lack of time for people on soft money to curate dark data. (Anonymous Respondent, 2016 pre-workshop survey)
\end{quote}

Some of the proposed solutions relevant to this effort include development of centralized curation resources where costs could be amortized across projects and time, with skilled expert staff who can apply similar solutions across many projects and mechanisms to assure academic credit, including embargoes, data creator controls to access, and creation of DOIs for data:

\begin{quote}
The key issue is funding for initial ingestion and perpetual curation. There is also a social issue associated with private observatories who may need to be convinced to release their proprietary data products; funding could help here, too.  Also, even major observatories have stores of dark data, e.g., the NOAO "save-the-bits" tapes. Interacting with these stakeholders will perhaps be more productive than scavenging piecemeal data sets from individual investigators. (Anonymous Respondent, 2016 pre-workshop survey)
\end{quote}

\subsubsection{Take advantage of CyVerse infrastructure and longevity of University of Arizona}

Leveraging dedicated CyVerse resources to support astronomical research and associated software requires identifying and incorporating corresponding software and metadata for astronomical datasets. Through CyVerse, it will also be possible to study user participation through established metrics, as well as observing broader applications for the astronomy and astrophysics communities as Astrolabe develops. Astrolabe will take advantage of both the CyVerse infrastructure as a computing environment, as well as the astronomy activities on campus and the longevity and stability of the University of Arizona as an institution, to establish a secure and robust repository that focuses on the needs of its users. Overall, Astrolabe shall support user control of assets and open access to research data, with a commitment to sustainability and education, to successfully navigate and avoid the following data curation challenges:

\begin{quote}
Inadequate human resources to process and document [data]. Loss of institutional memory of data properties. Lack of academic or professional credit for work on dark data that won't yield peer-reviewed publications. Unreadable legacy media. Scattered locations of the data. Loss of software needed to make sense of old data. (Anonymous Respondent, 2015 pre-workshop survey)
\end{quote}

However, as indicated by \citet{Sands2014}, disciplinary expertise and buy-in are essential to support curation efforts. As a result, the interdisciplinary Astrolabe project team should be comprised of individuals familiar with techniques for astronomy data mining and visualization as well as existing resources typically utilized by astronomers.

\subsubsection{Obtain community buy-in and manage expectations}

To solicit buy-in and to ensure that the systems being built meet community needs, an advisory board is critical for the project. Some board members may embrace active roles helping to incorporate their own datasets and advise project staff. Astrolabe must participate in a complex data space including many ongoing and completed observation missions as well as data service projects and virtual observatories, where:

\begin{quote}
The community doesn't acknowledge or reward archiving your data except for a few large catalogs (and even those aren't rewarded enough) ... There aren't standards or tools for how to get your data into a archivable form that are easily accessible to the average small-team researcher. There aren't tools that allow other people to access data or even published tables quickly and easily.  The VO effort spent a lot of time on standards, and basically produced things that are only of interest to large projects because the burden of making your dataset or tool compliant with a VO standard is more than any individual has time to deal with, especially when they aren't getting paid to do it. (Anonymous Respondent, 2016 pre-workshop survey)
\end{quote}

As envisioned, the Astrolabe project is essentially attempting to perform cultural engineering, changing the publishing habits of scientists to meet new open access requirements:

\begin{quote}
Technical issues are generally solved or solvable, but include data compression, accurate timestamps, attaching spatial and spectral coordinates, getting the community to cooperate on common standards, adopting underlying technologies likely to persist indefinitely, etc. (Anonymous Respondent, 2016 pre-workshop survey)
\end{quote}

Chief resources to coordinate with include: the Harvard-Smithsonian Center for Astrophysics, ADS, and other repositories such as Dataverse and Zenodo. Furthermore, the UAT can be used in Astrolabe, providing a taxonomy for semantic enrichment \footnote{http://astrothesaurus.org/}. In order to connect to existing resources such as ADS it is necessary to establish a presence at key conferences, and to meet with representatives of related projects, with the goal of educating the broader community about the new CyVerse resources as well as obtaining feedback from the community of users about their desires for the project's next steps through extramurally funded development.  Finally, education is a critical element for the viability and long-term sustainability of Astrolabe. This includes two main populations: astronomers who deposit data, and students in astronomy and information science who need to learn data science techniques. The Astrolabe team must work with authors of data to ensure that the data are in standard format and with appropriate metadata, and to develop best practice guidelines for astronomy data.

\subsubsection{Focus on low-hanging fruit}

Workshop participants identified two obvious pieces of ``low-hanging fruit" as a niche opportunity for Astrolabe to provide valuable services to the astronomical community: dark data and other orphan datasets not curated elsewhere, and data associated with authors of articles published in AAS journals:

\begin{quote}
As a point of discussion, the highest priority could be assigned to the data underlying refereed publications, along with medium-size datasets with uniform calibration that do not currently go to specific ground-based archives. (Anonymous Respondent, 2015 pre-workshop survey)
\end{quote}

As noted by \citet{Henneken2011}, publications based on datasets are essentially expressions of data. Journal publishers are innovating new methods of digital publication that provide rich scientific data beneath a text publication itself. Furthermore, as citation rates appear to be higher for publications that contain links to referenced data, participating in data citation and persistent linking is an important objective of Astrolabe \citep{Accomazzi2011a,Accomazzi2011b}. Additional synergistic activities include working with the WWT\footnote{http://worldwidetelescope.org}, a community-based tool for research, publishing, education and public outreach. CyVerse represents an outstanding partner, and the CyVerse software stacks are very similar to others in use across the physical sciences community. Additionally, the AAS is working to develop a community-based software discovery portal for astronomy with robust developer workflows, unique identifiers, software citation, search and developer credit \citep{Henneken2017}. Astrolabe should pilot an instance of this discovery portal in the data repository. Astrolabe is not limited to data that can be represented in WWT but can store any data format for visualization with packages such as Glue \citep{Beaumont2015}.

\subsubsection{Develop a follow-on workshop}

Workshop participants advised holding a subsequent workshop to connect community members as the system develops, to include both an education component and continued assessment of community needs. This follow-on workshop was held in 2016 and is described below. Additional future workshops should include data carpentry\footnote{http://www.datacarpentry.org/} plus hackathons, particularly targeting researchers with relevant datasets.

\subsection{2016 Astrolabe Workshop Outcomes}

2016 Astrolabe follow-on workshop participants indicated the following important areas of focus, which are now priorities for ongoing system development.

\subsubsection{Data complexity, heterogeneity and physical format of dark data}

\begin{quote}
It is rare that literally retracing someone else's data analysis path is going to be worthwhile. What we have is the issue of reusability. Our data should often be useful for multiple projects, but access problems make that more difficult than it needs to be. (Anonymous Respondent, 2016 pre-workshop survey)
\end{quote}

Historical data can be made accessible through Astrolabe. Datasets stored on magnetic tapes at NOAO and the Lunar and Planetary Laboratory (LPL) would likely be of interest to the community. Promising funding sources for these efforts were identified by UA Library participants, specifically for data preservation in the U.S. national library space. Similar efforts at the Harvard-Smithsonian Center for Astrophysics involving astronomical plates represent a possible partnership for archival expertise\footnote{https://platestacks.cfa.harvard.edu/}, and additional funding opportunities exist for such efforts. These data represent an ideal opportunity for citizen science projects as well. 

Larger datasets can also fall into the category of dark data, with complexity of the data leading to inaccessibility for reuse and research transparency: 

\begin{quote}
There are also very complex data sets that are difficult to host, think mini-SDSS type databases that include spectra, images, figures and tables all integrated by a search engine.  How do you integrate something like that? (Anonymous Respondent, 2015 pre-workshop survey)
\end{quote}

Furthermore, individual missions are able to specify the format of their data to create a certain amount of consistency. Individual researchers, however, structure their data to individual needs:

\begin{quote}
Astronomers don't want to properly separate final data products from intermediate files in their ``working directory".  A final measurement may come from a file having some unique pattern in the name, but is mixed in with all the files leading to the creation of that spectrum and only the astronomer understands the naming scheme. (Anonymous Respondent, 2015 pre-workshop survey)
\end{quote}

This means that a repository designed to support dark data must have staff and cyberinfrastructure that can manage this difficulty without putting undue burden on the individual researchers:

\begin{quote}
How to make such a heterogeneous set of data searchable?  Extractable? (Anonymous Respondent, 2015 pre-workshop survey)
\end{quote}

If cyberinfrastructure is shared across projects it can be more cost effective and long-lived than a scenario in which each researcher builds infrastructure for their own instruments or projects. It is also fundamentally important that any new service dealing with uncurated data should coordinate with other repositories to maximize resources and expertise and avoid redundant efforts:

\begin{quote}
There are other efforts to create a data hub on a national scale.  I am aware of a group at the National Center for Supercomputing Applications and the National Data Service who are working to build a massive data hub for storage of datasets pan-science.  To date, they have a working prototype involving manipulation of data in situ on their servers to avoid too many large data transfers of massive datasets.  It may be worthwhile to communicate with them to see if a shared infrastructure is warranted.  Two autonomous efforts by the both of us may not be as strong as one where we work together. (Anonymous Respondent, 2015 pre-workshop survey)
\end{quote}

\subsubsection{Author websites archiving data}

Authors frequently link to data on personal websites from the literature, and journals do not enforce a guarantee that these websites will be long-lived:

\begin{quote}
Often, collaborations that are medium size (multi year but much smaller than SDSS) create databases to serve their own uses and the public.  These generally fail some years after the collaboration is done, because the database software gets out of date or a server is upgraded, etc, and no one is getting paid any longer to make sure it's available. Maybe this is an application for the proposed data hub ... I think you would do a great service if you could make published data easily usable and cross-referenceable. Dark data is interesting but re-extracting quantities from it is unattractive, especially if they have already been published. (Anonymous Respondent, 2016 pre-workshop survey)
\end{quote}

Astrolabe could provide authors with a website tool through CyVerse, ensuring the integrity of links in published literature. This would allow authors to freeze a site at publication to preserve a record of the research process. Several existing web pages should be ingested as a proof of concept. Search tools are necessary to locate files across collections. Templates could be created through CyVerse, and users could adopt templates for the appropriate data structure, including key presentation and descriptive information. Metrics are needed to demonstrate usage. 

\subsubsection{Time domain and serendipitous data cases}

A key opportunity for Astrolabe is supporting time domain astronomy, including follow up to Large Synoptic Survey Telescope (LSST) observations, and bringing together datasets and researchers from multiple telescopes, if possible:

\begin{quote}
Re-observation of a very faint target can be expensive in telescope time.  Missing a point in a time domain sequence can impact the interpretation, or require negotiation with a person that the investigator did not necessarily intend to be a co-author.  (Anonymous Respondent, 2015 pre-workshop survey)
\end{quote}

\citet{Darch2017} show that the LSST project faces substantial uncertainty regarding future stakeholder requirements, including users' research practices, required tools and expertise, and the ability of other instruments to join the network for follow-up observations. Open source development places a burden on the community to create adaptive infrastructures to effectively interface with LSST data. Digital libraries such as Astrolabe must therefore anticipate and respond to change, promote standardization, develop infrastructures that empower users to adapt and reconfigure as needed, and adopt open source policies.

Another potential source of serendipitous discovery is historical data on nearly obsolete, institutional and individual removable media (as discussed above), which could be converted to current formats and ingested into Astrolabe. Historical data represent an opportunity for citizen science and educational opportunities. The amateur astronomy community is also a potential target for data. It is recommended that Astrolabe start with a minimum of 100 Terabytes of storage space, planning to scale up to ten Petabytes over ten years. The system should include the following features: support for a visualization interface; minting DOIs; checksum to detect errors in file transfer; licensing; capability to upload notes and other historical metadata related to data; public/private options and related policies; support for building tools; and capturing provenance of data. All of these features can be supported by CyVerse technology. To build a sustainable framework, Astrolabe should consider federation with other systems, and highlight costs saved by utilizing CyVerse in proposals to NSF and other agencies, including private foundations.

\subsubsection{Searching the literature for dark data}

Text mining could locate references to dark or potentially-dark data in the literature. Challenges associated with curation of these data provoke the following questions:

\begin{quote}
How do you make "dark data" findable? How to improve the "citability" of "dark data"? Does it make sense to write "data release" papers for "dark data", which would improve findability in publication-centric discovery services like the ADS? How do you measure the impact of "dark data" or at least whether it is being used? (Anonymous Respondent, 2016 pre-workshop survey)
\end{quote}

With the availability of a convenient resource for archival such as Astrolabe, authors could be incentivized to share data. High impact papers and publications within the past ten years could be targeted first. The ADS could help with relevant expertise and community support, and text mining to locate papers without data links. Candidate data associated with an existing publication would advertise for Astrolabe and provide growth and overall sustainability for the project. 

The standards of the Virtual Observatory support searching across repositories, but also lead to challenges associated with data heterogeneity:

\begin{quote}
Contrast with the Virtual Observatory paradigm that cooperates on standards but has no central structure to ensure continuity. (Anonymous Respondent, 2016 pre-workshop survey)
\end{quote}

\section{Discussion} \label{sec:discussion}
	
Workshop and survey participants overwhelmingly confirmed that dark and legacy data - along with some data associated with current small- to medium-sized projects - in astronomy represent a valuable target for curation efforts. The issues associated with dark or at-risk data can be organized into three categories, which were discussed at length in both Astrolabe workshops and are summarized here: 1) The science use cases for bringing dark data to light; 2) Where to find useful dark data; and 3) Governance, sustainability and buy-in.

\subsection{Science Use Cases}

	Any repository needs to support new science as a core justification for its existence. While it is impossible to determine what data will be valuable in the future, there are indications of what might be most valuable. As discussed extensively in section \ref{sec:outcomes} above, Astrolabe is scientifically justified and should increase communication of prior work for integration with new data, reanalysis under new methods with new data, verifiability and replicability. One example is the ability to revisit prior observations when a celestial event is recorded with LSST, which was highlighted in Astrolabe workshops as a strong motivator to look back in time to old observations. LSST will allow researchers to identify events across large sections of the sky, rekindling interest in how objects of study were behaving in the past. Availability of historical data would contribute to ongoing study of phenomena that change over time. Proper motion benchmarks represent another important application for Astrolabe archival data.

\subsection{Finding Dark Data}

	Many workshop participants and Astrolabe Advisory Board members have provided pointers for locating valuable data in danger of being lost. This includes not only older data, but also current or recent projects with uncertain sustainability. Three main types of old data sources were identified. The first is large data collections that were intentionally created for preservation purposes but now are underutilized because of data format obsolescence. The second is data associated with publications. The third is data from instruments that are no longer operating. Some of these data may have an institutional home, but because of time and costs, have not been migrated to more accessible formats and platforms. One of many examples is NOAO's tape data archive. We are currently developing plans to move 100 Terabytes of exabyte tapes and other media to more accessible formats. Our workshop participants have further identified many facilities and individuals with boxes of media of different types in storage rooms and filing cabinets worldwide. Previous work of this type includes the Digital Access to a Sky Century @ Harvard (DASCH) project that aims to scan the majority of the Astronomical Photographic Plate Collection's 500,000 glass plate negatives and produce full photometry results for the entire sky \citep{Simcoe2006}.

Many publishers are now making some data associated with journal articles available online. However, this is not universal and few such records exist for older publications. Since all historical data cannot be collected, participants suggested focusing on: high impact articles under the reasonable assumption that these data would be of greatest use; recent publications because the probability of recovering the data from the authors is highest; and finally, identification of outdated or broken data references and links in existing publications. While the three categories described here only address a small portion of the data that participants identified as important, Astrolabe is working with the AAS in each of these areas. We are currently searching the last ten years of publications for interesting data, with plans in place to create semi-automated methods of identifying data references, beginning with locating broken hyperlinks in publications. There are linguistic references to data as well that do not always refer to existing collections, but these are more difficult to identify in large collections of literature.  For example, a paper may refer to a dataset from a particular observing instrument that was operational at the time of publication while the instrument's data archive is no longer available online. We plan to develop pattern matching and machine learning tools to help identify these references.

Current projects are also sources of dark data, and this category overlaps substantially with new funding models. We are unable to prevent new data from becoming dark unless we solve the financial problems associated with storing, processing and disseminating data. The same social and technical influences that cause older data to become dark can also impact newer data. For example, three-year agency-funded projects can produce data streams that are difficult to maintain after the funding period expires. The widespread requirements for data management plans are not accompanied by additional funding for data management beyond the duration of a project. Smaller instruments and missions may not be archived in a long-lived repository even if all data collected are not analyzed by the end of the funded project. Astrolabe is developing partnerships and new research proposals with current instrumentation projects to construct efficient workflows for new data and digitization of old data that are identified as potentially useful. 

\subsection{Governance, Sustainability and Buy-in}

	Finally, the community has pointed out the critical element of governance, sustainability and buy-in. We have formed an advisory board of experienced community members to oversee development and funding. We have identified several principles that capture some of the important points discussed above. The Astrolabe repository must be open to all researchers, facilitating efficient access to data. It must have a low cost of operation. It must reuse or improve appropriate technologies such as the VO initiative and pan-science initiatives like Cyverse and JetStream rather than build new technologies wherever possible. It must be able to provide services that will encourage new projects to buy-in to meet their own goals and assure the long-term operation of facilities. The repository must be more than a passive storage location for data. Astrolabe needs community involvement to install and process data and metadata. Each of the two workshops presented here directed the Astrolabe project team towards key uncurated or at-risk datasets and strategies for locating and curating these data. Some of the ideas can be incorporated into existing repositories and many are being built into the Astrolabe repository.

One potential strategy for sustainability is to pull together resources from many data-gathering projects staggered in time. Another partial solution can be found from an analogy to publication of scholarly papers. A percentage of the overhead costs from research grants received by university researchers is designated for supporting institutional journal subscriptions, which then supports the editing and publishing of journals. Libraries maintain digital and paper collections also supported in part by overhead on research grants. Data management might in part be supported by overhead on grants, but this model for peer-reviewed publication has resulted in economic problems with rising journal subscription costs making access prohibitively expensive for individuals and institutions. The open access movement now directs some of the costs away from subscriptions, and this model might also be applied to data with a data deposit fee. Such a fee could be built into grant proposals, but in the end would need to cover the cost for curation throughout the entire data lifecycle. If Moore's law holds for data storage, perhaps some current data costs can be covered with future projects that would require much larger datasets. The Dryad repository for life sciences is an example of a successful business model for nonprofit governance through a membership organization\footnote{datadryad.org}.

\section{Conclusions and Future Directions} \label{sec:conclusions}

Data creators and researchers typically require motivation and incentive to share data, such as a clear vision for future use of the data in a broader context and professional credit for reuse. There must be low barriers to the technical process of actually putting data into a repository. Furthermore, researchers must maintain a certain level of control of their data over time to ensure proper attribution and quality control. We identified factors that impact the willingness and ability to share and use scientific data in astronomy, considering that data are only as useful as the software used to analyze and visualize the data.  Beyond simply creating another data repository, participants in the two workshops discussed here identified advantages to using \deleted{a}cyberinfrastructure that could facilitate not only data deposition and subsequent discovery, but also significant processing of that data, resulting in the creation of Astrolabe as a community resource that is now available for widespread use by astronomers through CyVerse.
	
New instrumentation and accelerated data flows in astronomy are posing a challenge to traditional approaches to experimentation and computing. For small- and medium-sized projects, there are no clear options for long-term sustainability in new data management plan requirements \citep{Hanisch2017}. CyVerse and related platforms such as JetStream\footnote{https://jetstream-cloud.org/}\citep{Stewart2015, Towns2014} exemplify the type of beneficial curation and computational environment envisioned by workshop participants. This cloud-based approach allows for flexible computing where one to hundreds of processors can be applied to a problem.  We will incorporate a community developed controlled vocabulary (i.e., the UAT) to enhance findability of datasets. These data will be processed in CyVerse and then integrated with the WorldWide Telescope visualization tool by the end of our current National Science Foundation grant in 2019. Further supporting discoverability, we have developed capabilities for FITS header extraction and semi-automated population of metadata templates configured according to the IVOA ObsCore, FITS, and WCS standards. Furthermore, when datasets in Astrolabe are determined to be static and ready for public access, CyVerse is capable of minting DOIs with a DataCite\footnote{https://www.datacite.org/} template, which permits linking to Astrolabe from ADS and connecting data to literature.

This paper has described the data curation recommendations of participants in two community workshops, and we have provided details about the ongoing development of the Astrolabe repository system to enable curation of potentially-valuable dark or at-risk data in astronomy. Overall, we wish to expand the capacity of the system to increase ease of use, data discoverability, convenient access to data manipulation, and incorporating additional data types, including not only data being generated from instruments aimed into space, but also remote sensing instruments pointed back down toward earth, such as Organization of Biological Field Stations site data that could also be visualized in WWT. Shared infrastructure presents curation challenges that must be navigated with the participation of users. Establishing an active Astrolabe user community will help us to identify both domain-specific computational tools and data for the shared environment, and also to exploit domain-spanning cyberinfrastructure, enabling grand challenge research and serendipitous discovery in multiple domains.
    
    \acknowledgments

We gratefully acknowledge the support of the National Science Foundation (grant number 1642446, P.B. Heidorn, PI; J. Steffen, Co-PI), the University of Arizona (2015 “Start for Success” grant and 2016 “Accelerate for Success” grants, P.B. Heidorn, PI; D. Zaritsky, Co-PI), the American Astronomical Society, and CyVerse. We thank our additional collaborators at the University of Arizona, including the UA Department of Astronomy and the University Libraries, as well as workshop organizers and participants for their time, engagement, and expertise. We especially thank Phil Rosenfield for his thorough feedback on the manuscript, along with our anonymous reviewer.


\begin{thebibliography}{}

\bibitem[Accomazzi(2011)]{Accomazzi2011a} Accomazzi, A. 2011, in Future professional communication in astronomy II (Cambridge: FPCA-II), 135

\bibitem[Accomazzi \& Dave(2011)]{Accomazzi2011b} Accomazzi, A. \& Dave, R. 2011, Semantic interlinking of resources in the virtual observatory era, arXiv preprint arXiv:1103.5958 

\bibitem[Accomazzi et al.(2011)]{Accomazzi2016} Accomazzi, A., Kurtz, M. J., Henneken, E. A., et al. 2016, Aggregation and Linking of Observational Metadata in the ADS, arXiv preprint arXiv:1601.07858 

\bibitem[Anderson(2004)]{Anderson2004} Anderson, C. 2004, The long tail, Wired magazine, 12, 170 

\bibitem[Atkins(2003)]{Atkins2003} Atkins, D. 2003, Revolutionizing science and engineering through cyberinfrastructure: Report of the National Science Foundation blue-ribbon advisory panel on cyberinfrastructure 

\bibitem[Beaumont et al.(2015)]{Beaumont2015} Beaumont, C., Goodman, A. \& Greenfield, P. 2015, Astronomical Data Analysis Software an Systems XXIV (ADASS XXIV) (Sydney: ADASS), 101 

\bibitem[Borgman(2015)]{Borgman2015} Borgman, C. L. 2015, Big data, little data, no data: Scholarship in the networked world (Cambridge: MIT press) 

\bibitem[Borgman et al.(2016)]{Borgman2016} Borgman, C. L., Golshan, M. S., Sands, A. E., et al. 2016, Data management in the long tail: Science, software, and service, International Journal of Digital Curation, 11, 128 

\bibitem[Brunner et al.(2001)]{Brunner2001} Brunner, R. J., Djorgovski, S. G. \& Szalay, A. S. 2001, Virtual Observatories of the Future (Pasadena: ASP) 

\bibitem[Choudhury et al.(2008)]{Choudhury2008} Choudhury, S., DiLauro, T., Szalay, A., et al. 2008, Digital data preservation for scholarly publications in astronomy, International Journal of Digital Curation, 2, 20 

\bibitem[Conrad et al.(2017)]{Conrad2017} Conrad, A. S., Rasmus, H. \& Svendsen, M. 2017, Reuse for Research: Curating Astrophysical Datasets for Future Researchers, International Journal of Digital Curation, 12, 37

\bibitem[Cragin et al.(2010)]{Cragin2010} Cragin, M. H., Palmer, C. L., Carlson, J. R., et al. 2010, Data sharing, small science and institutional repositories, Philos.Trans.A.Math.Phys.Eng.Sci., 368, 4023 

\bibitem[CyVerse(2017)]{CyVerse2017} CyVerse. 2017, Transforming science through data-driven discovery: CyVerse services overview, 2017 

\bibitem[Darch \& Sands(2017)]{Darch2017} Darch, P. T. \& Sands, A. E. 2017, Digital Libraries (JCDL), 2017 ACM/IEEE Joint Conference Anonymous (Toronto: IEEE), 1 

\bibitem[Edwards et al.(2011)]{Edwards2011} Edwards, P. N., Mayernik, M. S., Batcheller, A. L., et al. 2011, Science friction: Data, metadata, and collaboration, Soc.Stud.Sci., 41, 667 

\bibitem[Folk et al.(2011)]{Folk2011} Folk, M., Heber, G., Koziol, Q., et al. 2011, Proceedings of the EDBT/ICDT 2011 Workshop on Array Databases (Uppsala: ACM), 36 

\bibitem[Frey \& Accomazzi(2018)]{Frey2018} Frey, K. \& Accomazzi, A. 2018, The Unified Astronomy Thesaurus: Semantic metadata for astronomy and astrophysics, Astrophysical Journal Supplement Series (In Press)

\bibitem[Greenfield et al.(2015)]{Greenfield2015} Greenfield, P., Droettboom, M. \& Bray, E. 2015, ASDF: A new data format for astronomy, Astronomy and Computing, 12, 240 

\bibitem[Grosbol(1988)]{Grosbol1988} Grosbol, P. 1988, The FITS Data Format, Bulletin d'Information du Centre de Donnees Stellaires, 35, 7 

\bibitem[Hanisch et al.(2017)]{Hanisch2017} Hanisch, R., Hildreth, M., McEwen, L., et al. 2017, MPS Open Data Report 

\bibitem[Hanisch et al.(2007)]{Hanisch2007} Hanisch, R., Steffen, J., Choudhury, S., et al. 2007, Library and Information Services in Astronomy V (Cambridge: Wolbach Library), 29 

\bibitem[Heidorn(2008)]{Heidorn2008} Heidorn, P. B. 2008, Shedding light on the dark data in the long tail of science, Library Trends, 57, 280 

\bibitem[Henneken et al.(2017)]{Henneken2017} Henneken, E., Accomazzi, A., Blanco-Cuaresma, S., et al. 2017, Asclepias - Capturing Software Citations in Astronomy 

\bibitem[Henneken et al.(2009)]{Henneken2009} Henneken, E., Accomazzi, A., Kurtz, M., et al. 2009, Astronomical Data Analysis Software and Systems XVIII (Québec City: ADASS), 384 

\bibitem[Henneken(2015)]{Henneken2015} Henneken, E. 2015, Unlocking and sharing data in astronomy, Bulletin of the Association for Information Science and Technology, 41, 40 

\bibitem[Henneken \& Accomazzi (2011)]{Henneken2011} Henneken, E. A. \& Accomazzi, A. 2011, Linking to data-effect on citation rates in astronomy, arXiv preprint arXiv:1111.3618 

\bibitem[Keeney et al.(2001)]{Keeney2001} Keeney, S., Hasson, F. \& McKenna, H. P. 2001, A critical review of the Delphi technique as a research methodology for nursing, Int.J.Nurs.Stud., 38, 195 

\bibitem[Lenhardt et al.(2016)]{Lenhardt2016} Lenhardt, W. C., Conway, M., Scott, E., et al. 2016, High Performance Extreme Computing Conference (HPEC), 2016 IEEE (Waltham: IEEE), 1 

\bibitem[McCray(2004)]{McCray2004} McCray, W. P. 2004, Giant telescopes: Astronomical ambition and the promise of technology (: Harvard University Press) 

\bibitem[McCray(2014)]{McCray2014} McCray, W. P. 2014, How astronomers digitized the sky, Technology and Culture, 55, 908 

\bibitem[Michener et al.(1997)]{Michener1997} Michener, W. K., Brunt, J. W., Helly, J. J., et al. 1997, Nongeospatial metadata for the ecological sciences, Ecol.Appl., 7, 330 

\bibitem[National Academies (2016)]{National2016} National Academies of Sciences, Engineering, and Medicine. 2016, Future directions for NSF advanced computing infrastructure to support US science and engineering in 2017-2020 (: National Academies Press) 

\bibitem[Novacescu et al.(2018)]{Novacescu2018} Novacescu, et al. 2018, A model for data citation in astronomical research using Digital Object Identifiers (DOIs), Astrophysical Journal Supplement Series (In Press)

\bibitem[NSF(2015)]{NSF2015} NSF. 2015, Public Access Plan: Today's Data, Tomorrow's Discoveries: Increasing Access to the Results of Research Funded by the National Science Foundation 

\bibitem[Open Science Collaboration(2012)]{Open2012} Open Science Collaboration. 2012, An open, large-scale, collaborative effort to estimate the reproducibility of psychological science, Perspectives on Psychological Science, 7, 657 

\bibitem[Rosenfield(2018)]{Rosenfield2018} Rosenfield, P. 2018, AAS WorldWide Telescope: Seamless, cross-platform data visualization engine for astronomy research, education, and democratizing data, Astrophysical Journal Supplement Series (In Press)

\bibitem[Sands et al.(2014)]{Sands2014} Sands, A. E., Borgman, C. L., Traweek, S., et al. 2014, We’re working on it: transferring the sloan digital sky survey from laboratory to library, International Journal of Digital Curation, 9, 98 

\bibitem[Simcoe et al.(2006)]{Simcoe2006} Simcoe, R., Grindlay, J., Los, E., et al. 2006, An ultrahigh-speed digitizer for the Harvard College Observatory astronomical plates, arXiv preprint astro-ph/0610351 

\bibitem[Stahlman et al.(2018)]{Stahlman2018} Stahlman, G. S., Heidorn, P. B. \& Steffen, J. 2018, Proceedings of Library and Information Services in Astronomy (LISA) VIII (Strasbourg: CDS) (In Press)

\bibitem[Stewart et al.(2015)]{Stewart2015} Stewart, C. A., Cockerill, T. M., Foster, I., et al. 2015, Jetstream: A self-provisoned, scalable science and engineering cloud environment 

\bibitem[Towns et al.(2014)]{Towns2014} Towns, J., Cockerill, T., Dahan, M., et al. 2014, XSEDE: accelerating scientific discovery, Computing in Science \& Engineering, 16, 62

\bibitem[White et al.(2009)]{White2009} White, R., Accomazzi, A., Berriman, G., et al. 2009, The High Impact of Astronomical Data Archives (Astro2010: The Astronomy and Astrophysics Decadal Survey-Position Paper 64)

\end{thebibliography}
\end{document}